# US College Net Price Prediction Comparing ML Regression Models

Zalak Patel, Ayushi Porwal, Kajal Bhandare, Jongwook Woo

Department of Information Systems, California State University, Los Angeles

{zpatel6, aporwal, kbhanda3, jwoo5}@calstatela.edu

**Abstract**
This paper will illustrate the usage of Machine Learning algorithms on US College Scorecard datasets. For this paper, we will use our knowledge, research, and development of a predictive model to compare the results of all the models and predict the public and private net prices. This paper focuses on analyzing US College Scorecard data from data published on government websites.

Our goal is to use four machine learning regression models to develop a predictive model to forecast the equitable net cost for every college, encompassing both public institutions and private, whether for-profit or nonprofit

## 1. Introduction

As a student or their parent, do we feel that the net price we are paying for college is worth it? Whether we've just started to create the list of potential colleges or we're in the process of narrowing down the options, one of the key considerations for many of us is, well, money. We're all aware that college can come with a hefty price tag, but what may surprise us is that the cost isn't the same for everyone. It varies based on multiple aspects such as colleges and universities across the US, their demographics, graduation rates, tuition costs, student loans, and much more, including average salaries after graduation. College should be within reach of all students. So have built this model as an initiative to support students to find affordable options and make potential decisions for future careers.

We will use data from the US College Scorecard to predict college prices. This data includes details about the schools, admission requirements, and financial aid. By using a method called Random Forest Regressor, we aim to estimate fair college prices accurately. This helps us understand how different factors affect college costs. Ultimately, it helps students and families make informed decisions about college affordability.

## 2. Dataset Details

The dataset used in this paper is taken from the US Department of Education website. Data is up to May 06, 2024. To achieve the limit of 2GB, we concatenated 12 Excels starting from the year 2010 to the year 2022 with 172 columns and about 3 M rows. Each row in the dataset represents the college from a distinct year and its price.

**Dataset URL**: https://collegescorecard.ed.gov/data/
A few of the columns from our dataset are mentioned below:
- NPT4_PUB: This field will consist of the annual net price of public colleges.
- NPT4_PRIV: This field will consist of the annual net price of private colleges.
- COSTT4_A: This column represents the average cost of attendance for undergraduate students at Title IV institutions. The 'A' stands for 'all' undergraduate students.
- CONTROL: It will categorize colleges and universities based on whether they are public, private nonprofit, or private for-profit.

## 3. Paper Scope

As part of this study, we utilized PySpark, machine learning regression models to predict the net prices of both private and public universities. To enhance accuracy, we conducted data cleaning and feature engineering. Additionally, we fine-tuned hyperparameters and parameters through cross-validation and train-validation splits to enhance model performance. We compared various regression models to determine the one delivering the highest accuracy within the shortest timeframe. Furthermore, we assessed overfitting by evaluating the $R2$ on test data and comparing it with the training dataset.

## 4. Technical Specification

| # of CPU cores | 8 |
|---|---|
| # of Nodes | 5 (2 master nodes, 3 worker nodes) |
| CPU speed | 1995.309 MHz |
| Hadoop Version | 3.3.3 |
| Pyspark Version | 3.2.1 |
| Memory Size | 806.4 GB |

## 5. Related Work

Article published by Faster Capital on "The Power of Regression Analysis in Price Forecasting." We referred to this article to understand the importance of predicting prices and how regression models are best used to achieve accurate prediction [1].

Article published by the U.S Department of Education on "Updated College Scorecard Will Help Students Find High Value Postsecondary Programs" with this article we were able to understand the importance of the dataset on how it will help families, students, and educators make data-informed decisions when choosing a college or university to attend as it an open-source website. It also guides us on the important columns like student debt and earnings from the National Student Loan Data System and the Internal Revenue Service (IRS), graduate school outcomes, and longer-term earnings by college that resulted in significant features for our study [2].

Article published by Scientific Reports on "Using Machine Learning to Predict Student Retention," with this article we were able to understand feature selection and identify which features have a greater impact on the prediction of net price. In this article, the authors predicted the retention rate for universities using classification models, while we are predicting the net price for colleges using regression models [3].

Article published by Hassan Shahin on "US Colleges Scorecard Analysis". We referred to this article to understand the feature engineering, data manipulation pipeline, and selection of features and labels. In the referenced article, the author built and evaluated a random forest regression model using the Python library. In contrast, we have utilized additional feature engineering techniques, assessed feature importance, and conducted parameter tuning to build and evaluate the best-performing model from four different regression models (Random Forest, Linear Regression, Decision Tree, and Gradient-boosted tree regression) using Apache PySpark [4].

## 6. Background / Existing work

In our study, we have used four regression algorithms for College Net Price Prediction (Linear Regression, Random Forrest Regression, Gradient Boosted Tree Regression, and Decision Tree Regression models)

### 6.1 Regression

Regression serves as a supervised machine learning method aimed at forecasting continuous values. In the initial phase of our Net Price Prediction study, we leverage Regression models since the target variable, price, constitutes a continuous numeric entity. Employing four Regression algorithms—Decision Tree Regression, Gradient Boosted Tree Regression, Linear Regression, and Random Forest Regression—we utilize modules such as Tune Model Hyperparameters for enhanced model efficacy, Train Split Validator for assessing generalization, and Cross Validate for validating model performance. Additionally, the Permutation feature importance module is employed to iteratively eliminate less significant features. In the context of SparkML for price prediction, we deploy Gradient-Boosted Tree Regression, Decision Tree Regression, Linear Regression, and Random Forest Regression algorithms. These regression models are built upon prior lab work, specifically predicting 'arrival delay' in the flight dataset. Following a similar process of creating a pipeline for feature transformation and training regression models, we employ cross-validation and train split validation to determine optimal parameters and evaluation metrics such as Root Mean Square Error (RMSE) and Coefficient of Determination (R2) are utilized to assess model performance.

### 6.2 Flow chart and Architecture

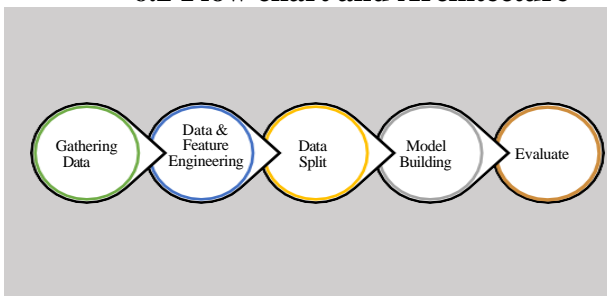

*Fig 1 Work Architecture*

**Gathering Data:** Input the data. That data can be structured or unstructured, originating from various sources and formats. The goal is to compile all relevant data necessary for analysis and model development.
**Data & Feature Engineering:** Frame the machine learning problem by identifying what needs to be predicted and determining the relevant features or observations required for making those predictions. This step involves cleaning, transforming, and selecting the most important features from the raw data to enhance model performance.
**Data Split:** Divide the dataset into training and testing subsets. The training data is used to build and train the model, while the test data is reserved for evaluating the model's performance on unseen data.
**Model Building:** Select an appropriate model based on the problem at hand and implement it using the training dataset. Train the model by fitting it to the training data and adjusting its parameters to learn the underlying patterns and relationships.
**Evaluate:** Use the test dataset to make predictions and evaluate the model's performance. Assess the model's accuracy, RMSE, R2, precision, recall, and other relevant metrics to ensure it meets the desired performance criteria. This step may also involve comparing the results with other models or approaches to identify the best solution.

## 7. Machine Learning Models

In our study, we implemented various machine learning models to predict the net price for colleges. To achieve this, we utilized multiple datasets spanning from the year 2010 to 2022. These datasets were consolidated into a single file, which was then used to build our predictive models.

Big Data efficiently stores and processes massive datasets using methods like Hadoop, data mining, and machine learning. Machine learning is a diverse and dynamic field that is well-suited for handling big data. It transforms raw data into actionable insights by automating the data analysis process. It enables us to automate tasks, identify patterns, and make predictions, thus enhancing decision-making processes.

### 7.1 Linear Regression

Linear regression is a method used to model the relationship between a dependent variable (scalar response) and one or more independent variables (explanatory variables). When there is only one explanatory variable, it is known as simple linear regression. When there are multiple explanatory variables, it is called multiple linear regression.

In our study, we implemented Linear Regression to predict net prices. We employed Cross-Validation and Train-Test Split Validation techniques to ensure the model's reliability and generalizability. The dataset was divided into 70% training data and 30% test data.

We tuned the model's hyperparameters to identify the best-performing configuration, using a parameter grid (Param Grid) for systematic tuning. Cross-validation was crucial in mitigating overfitting and ensuring that the model performs well on unseen data.

As shown in Table 7.1 the Linear Regression Algorithm yielded the same RMSE and R2 values when using both Train Split and Cross Validation methods. The first model is trained using Train-Validation Split, while the second model is trained using Cross-Validation. The results indicate that the RMSE value is quite high at 3314.532, and the R2 value is low at 0.7700. Therefore, it is evident that this model is not suitable for accurate predictions.

| Linear Regression Model | TVS | CV |
|---|---|---|
| Coefficient of Determination (R2) | 0.7700 | 0.7700 |
| Root Mean Square Error (RMSE) | 3314.5325 | 3314.5325 |

*Table 7.1 Linear Regression Model Performance Metrics*

### 7.2 Decision Tree Regression

As discussed in Section 7.1, Linear Regression is not the optimal model for price prediction due to its high RMSE values and low R2 values.
Consequently, we employed the Decision Tree Regression Algorithm for better accuracy.

We applied both Cross-Validation and Train Validation Split methods for Decision Tree Regression, splitting the data into 70% training and 30% testing sets. Model hyperparameters were fine-tuned to identify the best-performing model. Cross-validation was utilized to generalize the model, with parameters defined using a parameter grid. Table 7.2 displays the evaluation results for Decision Tree Regression.

| Decision Tree Regression Model | TVS | CV |
| --- | --- | --- |
| Coefficient of Determination (R2) | 0.7657 | 0.7657 |
| Root Mean Square Error (RMSE) | 3414.3596 | 3414.3596 |

*Table 7.2 Decision Tree Regression Model Performance Metrics*

Table 7.2 indicates that the R2 value for both the Cross Validation model and the Train Split Validation model is 0.7657. Additionally, the RMSE values are identical for both methods. It is evident that the performance metrics for Linear Regression were slightly better than those for the Decision Tree model.

### 7.3 Gradient Boosted Tree Regression

Gradient Boosted Tree (GBT) is a machine learning technique used in regression that builds an ensemble of weak prediction models, typically decision trees. When decision trees are used as weak learners, the resulting algorithm is called a gradient-boosting tree.

We applied the GBT regression model using both Cross-Validation and Train Validation Split methods. The data was split into 70% training and 30% testing sets. Model hyperparameters were fine-tuned to identify the best-performing model, with parameters defined using a parameter grid. Table 7.3 below displays the evaluation results for GBT regression.

| Gradient Boosted Tree Regression Model | TVS | CV |
| --- | --- | --- |
| Coefficient of Determination (R2) | 0.8480 | 0.8475 |
| Root Mean Square Error (RMSE) | 2773.3521 | 2778.0047 |

*Table 7.3 Gradient Boost Regression Model Performance Metrics*

As shown in Table 7.3, the R2 value for the Train Validation Split model is 0.8480, which is slightly better than the Cross-Validation model at 0.8475. Additionally, the RMSE for the Train-Validation Split model is 2773.352, compared to 2778.004 for the Cross-Validation model. Overall, based on the RMSE and R2 values, the GBT regression model outperforms the previously discussed regression algorithms.

### 7.4 Random Forest Regression

Random forests are among the most versatile and easy-to-use supervised learning algorithms, suitable for both classification and regression tasks, though they are primarily used for classification. A random forest is composed of multiple decision trees, each built on a random subset of the data and features, which enhances model accuracy and robustness by averaging the predictions of individual trees. Random forests have diverse applications, including recommendation engines, image classification, and feature selection.

All the above-listed models are not the best model for Price prediction. The feature Importance module was used to determine the best features to use in the model and Cross-Validation helped generalize the model.

| Random Forest Regression Model | TVS | CV |
| --- | --- | --- |
| Coefficient of Determination (R2) | 0.8471 | 0.8447 |
| Root Mean Square Error (RMSE) | 2724.0135 | 2744.9912 |

*Table 7.4 Random Forest Model Performance Metrics*

As observed in Table 7.4, the R2 value for the Train-Validation Split model (0.8471) is slightly higher than that of the Cross-Validation model (0.8447). Additionally, the Train-validation model has a lower RMSE (2724.013) compared to the Cross-Validation model (2744.991). Overall, based on the RMSE and R2 values, the Random Forest Regression model demonstrates its effectiveness as the best fit for predicting Net Price.

We also evaluated the Random Forest regression model for overfitting. Since we concluded that this model is the best fit, it is important to ensure that it is not overfitted. To accomplish this, we assessed the model's performance using RMSE and R2 for both the training and testing data. Our analysis revealed that there was no significant disparity in accuracy between the training and test datasets. The model demonstrated consistent performance across both training and test datasets, suggesting that it was able to generalize well to unseen data. This alignment between the model's performance on training and testing data indicates that it did not exhibit signs of overfitting. Based on these findings, we confidently concluded that our Random Forest regression model was not overfitted.

### 8. Hyperparameter Tuning

Hyperparameter tuning is a crucial step in optimizing the performance of machine learning models. In this study, we focused on fine-tuning the hyperparameters of several models: Random Forest, Gradient Boosted Trees (GBT), Decision Trees (DT), and Linear Regression (LR). Our primary goal was to achieve the highest predictive accuracy for the net price prediction of USA colleges. Each of these models has its own set of hyperparameters that significantly influence their performance and behavior.

We tuned hyperparameters such as the number of trees (n_estimators), the maximum depth of each tree (max_depth), the maximum number of bins (maxBins), minimum information gain (minInfoGain), maximum number of iterations (maxIter), the regularization parameter (regParam), the elastic net mixing parameter(elasticNetParam), and whether to standardize the features (standardization) to evaluate the best accurate model.

In our approach, we employed both train-validation and cross-validation techniques to systematically explore the hyperparameter space and evaluate the model's performance. This thorough exploration ensured that the final models were not only accurate but also generalizable, capable of performing well on both seen and unseen data. By striking a balance between bias and variance through our chosen hyperparameter configurations, we effectively minimized overfitting while maximizing predictive performance.

The resulting hyperparameter configurations demonstrated a balanced trade-off between bias and variance, contributing to minimized overfitting and enhanced predictive accuracy. This comprehensive approach not only bolstered the accuracy of our models but also reinforced their stability and consistency across various datasets. Ultimately, it empowered us to generate robust and reliable predictions for the net price of colleges in the USA, thereby providing valuable insights for decision-making processes in the educational domain.

## 9. Model Comparison Table

| Algorithms | Results | Time Taken to fit the model |
|---|---|---|
| **Train Validation Split (TVS)** | | |
| Random Forest Regression | R2: 0.8471 RMSE: 2724.013 | 128.63 sec |
| Gradian Boosted Trees Regression | R2: 0.8480 RMSE: 2773.352 | 2167.57 sec |
| Decision Tree Regression | R2: 0.7657 RMSE: 3414.359 | 133.31 sec |
| Linear Regression | R2: 0.7700 RMSE: 3314.532 | 513.58 sec |
| **Cross Validation (CV)** | | |
| Random Forest Regression | R2: 0.8447 RMSE: 2744.991 | 293.09 sec |
| Gradian Boosted Trees Regression | R2: 0.8475 RMSE: 2778.004 | 5182.68 sec |
| Decision Tree Regression | R2: 0.7657 RMSE: 3414.359 | 318.01 sec |
| Linear Regression | R2: 0.7700 RMSE: 3314.532 | 1458.97 sec |

*Table 1: Comparison Table for Algorithms Performance*

Based on the results presented in Table 1, we can arrange various Regression Algorithms in the following order.
**Based on time:**
GBT> LR > DT> RF (GBT taking the most time, while RF taking the least)
**Based on accuracy:**
RF> GBT> LR> DT (RF having the best accuracy, while DT having the least)

Thus, we can conclude that the Random Forest model with train-validation split is the best fit, as it achieves the lowest Root Mean Square Error (RMSE) and the highest R2. Additionally, there is very little difference in R2 between the Random Forest and Gradient Boosted Trees (GBT) models with train-validation split, but the Random Forest model requires significantly less time to train compared to Gradient Boosted Trees (GBT).

## 10. Conclusion

This paper aims to identify the best-performing models for predicting the net price of both public and private institutes across the US. We conducted predictive analysis on a Hadoop- Spark Cluster using various regression algorithms, including Random Forest, Gradient Boosted Trees (GBT), Decision Trees (DT), and Linear Regression (LR).

Our experimental results, summarized in Table 1, reveal that the Random Forest Regression model achieved an RMSE of 2724.013 and an R2 value of 0.84. These metrics indicate that the Random Forest algorithm is the most effective model for net price prediction when compared to the other regression algorithms evaluated.

After analyzing the data and employing various predictive models, we can conclude that predicting net prices for US colleges is influenced by a variety of factors. Our models showed that factors like average cost of attendance, institution type, and in-state and out-of-state tuition fees are crucial in determining net prices. While our models provide valuable insights, it is essential to acknowledge the inherent complexity and variability in college pricing. Further research and fine-tuning of predictive algorithms are necessary to improve accuracy and adapt to changing trends in higher education costs.

In conclusion, our study underscores the significance of using advanced predictive models like Random Forest to estimate the net price of US colleges. While our findings contribute meaningful insights into the factors influencing college costs, ongoing research and model enhancements are crucial for adapting to the dynamic nature of higher education pricing. This approach will ultimately lead to more accurate and reliable predictions, aiding students, parents, and policymakers in making informed decisions regarding college affordability.

## 11. References

[1] The power of regression analysis in price forecasting - FasterCapital. Retrieved from https://fastercapital.com/content/The-Power-of-Regression-Analysis-in-Price-Forecasting.html.

[2] U.S. Department of Education & Posted by U.S. Department of Education. (2023, April 24). Retrieved from https://blog.ed.gov/2023/04/updated-college-scorecard-will-help-students-find-high-value-postsecondary-programs/

[3] Matz, S. C., Bukow, C. S., Peters, H., Deacons, C., Dinu, A., & Stachl, C. *Scientific Reports volume 13, Article number: 5705 (2023)*

[4] Hassan Shahin. US Colleges scorecard analysis. Kaggle. Published October 28, 2021. Retrieved from https://www.kaggle.com/code/hassanshahin/us-colleges-scorecard-analysis#Predictive-model.